\documentclass[aps,10pt, pre,twocolumn,superscriptaddress]{revtex4-1}
\usepackage{graphicx}
\usepackage{float}
\usepackage{amsmath}
\usepackage{mathtools} 
\usepackage{color}
\usepackage{afterpage}
\usepackage{xr}
\usepackage{braket}
\usepackage{makecell}
\usepackage{dsfont}

\usepackage{amssymb}
\usepackage{graphicx,color} 
\usepackage{bbm} 
\usepackage{multirow}
\usepackage{enumitem}
\usepackage{mathrsfs}
\usepackage{commath}
\usepackage{hyperref}
\usepackage{fontenc}
\usepackage[dvipsnames]{xcolor}
\hypersetup{colorlinks=true,urlcolor=blue,citecolor=blue,linkcolor=blue}
\usepackage{tabularx}

\usepackage[english]{babel}
\usepackage{float}
\usepackage{amsmath,amssymb}
\usepackage{mathtools} 
\usepackage{color}
\usepackage{xr}
\usepackage{graphicx}% Include figure files
\usepackage{dcolumn}% Align table columns on decimal point
\usepackage{bm}% bold math

\newcolumntype{Y}{>{\centering\arraybackslash}X}

\begin{document}

\title{Jamming, relaxation, and memory in a structureless glass former}

\author{Patrick Charbonneau}
\affiliation{Department of Chemistry, Duke University, Durham, North Carolina 27708}
\affiliation{Department of Physics, Duke University, Durham, North Carolina 27708}
\author{Peter K. Morse}
\thanks{Corresponding author}
\email{peter.k.morse@gmail.com}
\affiliation{Department of Chemistry, Duke University, Durham, North Carolina 27708}
\affiliation{Department of Chemistry, Princeton University, Princeton, NJ 08544}
\affiliation{Department of Physics, Princeton University, Princeton, NJ 08544}
\affiliation{Princeton Institute of Materials, Princeton University, Princeton, NJ 08544}

\date{\today}

\begin{abstract}
Real structural glasses form through various out-of-equilibrium processes, including temperature quenches, rapid compression, shear, and aging. Each of these processes should be formally understandable within the recently formulated dynamical mean-field theory of glasses, but many of the numerical tools needed to solve the relevant equations for sufficiently long timescales do not yet exist. Numerical simulations of structureless (and therefore mean-field--like) model glass formers can nevertheless aid searching for and understanding such solutions, thanks to their ability to disentangle structural from dimensional effects. 
We here study the infinite-range Mari-Kurchan model under simple non-equilibrium processes and compare the results with those from the random Lorentz gas [\textit{J. Phys. A: Math. Theor.} \textbf{55} 334001, (2022)], which are both mean-field--like and become formally equivalent in the limit of infinite spatial dimensions. Of particular interest are jamming from crunching and under instantaneous temperature quenches. The study allows for an algorithmic understanding of the jamming density and of its approach to the infinite-dimensional limit. The results provide important insight into the eventual solution of the dynamical mean-field theory, including onsets and anomalous relaxation, as well as into the various algorithmic schemes for jamming.
\end{abstract}

\maketitle

\section{Introduction}

The free energy landscape analogy has long been used to conceptually unify the structure, dynamics, and thermodynamics of supercooled liquids and glasses~\cite{anderson_lectures_1979, stillinger_topographic_1995, sastry_signatures_1998, debenedetti_supercooled_2001}. A roughening terrain, for instance, suggests why liquids should become sluggish upon cooling and then rigidify as one of many disordered metastable states. The intuitive nature of this representation has further found broad uses in fields ranging from ecology~\cite{altieri_properties_2021, altieri_effects_2022} to general relativity~\cite{facoetti_classical_2019}. For certain abstract models, the metaphor has even been shown to be exact~\cite{parisi_meanfield_2010, ros_complex_2019}. The equilibrium and out-of-equilibrium dynamics of fully-connected $p$-spin glass models, in particular, can be understood in terms of specific landscape features~\cite{parisi_infinite_1979, parisi_sequence_1980, panchenko_parisi_2013, kirkpatrick_dynamics_1987, cugliandolo_analytical_1993, bernaschi_strong_2020, folena_gradient_2021, folena_rethinking_2020}. The extension of the analogy to finite-size systems is also still actively pursued~\cite{ros_dynamical_2021, rizzo_path_2021, folena_equilibrium_2022}.

Following  the success of this approach, a similar theoretical program has been initiated for models that are conceptually closer to structural glasses. After theoretically solving for a number of static features of \emph{simple glasses} in the high-dimensional, $d\rightarrow\infty$ limit~\cite{charbonneau_glass_2017, parisi_theory_2020}, a formal dynamical mean-field theory (DMFT) description of these models has been formulated~\cite{maimbourg_solution_2016, szamel_simple_2017, agoritsas_outofequilibrium_2018, agoritsas_outofequilibrium_2019, agoritsas_outofequilibrium_2019a}. While symmetries of the resulting equations have been exploited to show the equivalence of response to various perturbations in both mean-field theory~\cite{agoritsas_meanfield_2021} and low dimensions~\cite{morse_direct_2021}, actual solutions of these equations are currently only available over relatively narrow equilibrium~\cite{manacorda_numerical_2020, baity-jesi_meanfield_2019} and low density~\cite{arnoulxdepirey_active_2021} conditions. The physical robustness of various finite-$d$ features of the landscape thus remain unclear. In particular, although the theoretical description of jammed hard spheres captures the remarkable criticality of systems as low as $d=2$~\cite{charbonneau_jamming_2015, torquato_jammed_2010, charbonneau_finitesize_2021}, key features of the out-of-equilibrium processes that bring systems to jamming remain somewhat murky. 

From a landscape perspective, one of the simplest out-of-equilibrium processes to consider is to instantaneously quench a liquid (or crunch hard spheres) equilibrated at a given temperature (or density) to its inherent state (IS)~\cite{speedy_random_1998}. However, the dynamics of that relaxation~\cite{folena_gradient_2021, stanifer_avalanche_2022, nishikawa_relaxation_2021, nishikawa_relaxation_2022, manacorda_gradient_2022, folena_weak_2023}, the IS algorithmic robustness~\cite{luding_much_2016, nishikawa_relaxation_2022}, and its memory of the initial density~\cite{ozawa_jamming_2012, berthier_equilibrium_2016, ozawa_exploring_2017, jin_jamming_2021, charbonneau_memory_2021} or temperature~\cite{sastry_signatures_1998, sastry_onset_2000, kamath_thermodynamic_2001, brumer_meanfield_2004, sastry_glassforming_2013, folena_rethinking_2020} are incompletely described. For hard spheres, even the lowest density at which jammed (disordered) states can be obtained remains unclear. Finite-$d$ results have been obtained~\cite{skoge_packing_2006, charbonneau_glass_2011, morse_geometric_2014, charbonneau_memory_2021}, and for some of these features, a reasonably robust phenomenology has been observed. Yet the significant influence of liquid structure in low $d$ systems over the accessible dimensional range~\cite{mangeat_quantitative_2016, charbonneau_dimensional_2022} obfuscates what the DMFT description might be. A different way to extrapolate to $d\rightarrow\infty$--and therefore to assess the robustness of the mean-field description as well as to extract physical insight from it--is to consider the $d$ convergence of models which exhibit a markedly simpler structure.

In this spirit, Manacorda and Zamponi have recently considered the jamming behavior of the simple random Lorentz gas (RLG)~\cite{manacorda_gradient_2022}, which is defined by the dynamics of an individual tracer particle moving in a system of fixed, uniformly distributed obstacles. By construction, the RLG model is formally equivalent to simple glasses in the limit $d\rightarrow\infty$, up to a trivial scaling factor (described below)~\cite{biroli_meanfield_2021, biroli_interplay_2021}. Given the single-particle nature of this model, its relaxation dynamics can be cleanly integrated, which currently enables numerical simulations up to $d=22$. Interestingly, this analysis has revealed unexpected features of the gradient dynamics. While unjammed systems reach an IS in a fairly short time, jammed systems appear to drift logarithmically with time toward an IS, thus calling into question the traditional algorithmic approach to jamming, which relies on fixing cutoffs for determining convergence. Surprisingly, the study obtained a significantly lower jamming density than had been previously reported for an analogous system~\cite[Fig.~9.2]{parisi_theory_2020}.

In order to assess the robustness of these findings, we consider a natural complement to the single-particle RLG, namely the many-body Mari-Kurchan (MK) model~\cite{mari_dynamical_2011}, which consists of $N$ particles interacting via random uncorrelated shifts. By construction, this model also exhibits a trivial structure in all $d$ and is described by the same DMFT equations as the RLG in the limit $d\rightarrow\infty$. Although preliminary work has been done to address the jamming behavior of the MK model~\cite{mari_dynamical_2011, parisi_theory_2020}, a comprehensive finite-size and finite-dimensional analysis is still lacking. In addition, the jamming algorithms used thus far are not straightforwardly amenable to theoretical analysis as they mix quenching with equilibrating. In this work, we provide such analysis and undertake a comparison to the relaxation dynamics of the RLG. While we validate many of the RLG findings, we find its estimate of the jamming density to be much lower than what any standard jamming algorithm can produce. This effort additionally allows us to characterize the IS memory and to conclusively determine that its onset transition, at which the IS starts depending on the initial condition, does not coincide with the dynamical transition, as it does in the pure $p$-spin model~\cite{cugliandolo_analytical_1993}.

The rest of this article is structured as follows. Section~\ref{sec:methods} presents the numerical MK planting techniques, gradient descent, and jamming algorithms. Section~\ref{sec:theoryHeader} summarizes known results about the threshold and dynamical transition in MK systems. Section~\ref{sec:relaxation} reports the relaxation dynamics in harmonic MK systems across spatial dimension, notably showing that they closely mirror those of the RLG at short times and that both converge to the same $d\rightarrow\infty$ dynamics at all times. We then discuss jamming in both soft and hard sphere MK models (Sec.~\ref{sec:hardsoft}). For the former, we specifically relate the findings to the RLG and to prior theoretical predictions. The existence of IS memory in both soft and hard sphere MK models is considered afterwards (Sec.~\ref{sec:memory}). Section~\ref{sec:conclusion} concludes by discussing the implications of these results for generic solutions of the DMFT and for more physical models of glasses in finite $d$.

\section{Computational Methods}
\label{sec:methods}

In this section, we describe the various computational approaches employed for obtaining IS of MK liquids. We first detail how to obtain efficiently equilibrated harmonic sphere configurations. We then present a gradient descent scheme for soft spheres and two different crunching algorithms for hard spheres. Each of these algorithms presents a different approach to the jamming line. 

\subsection{Planting for radially symmetric soft potentials}

Equilibrium MK configurations can generically be planted under any conditions, thus markedly reducing the computational cost of initializing systems. 
While hard spheres can be planted by simple rejection sampling~\cite{mari_dynamical_2011, charbonneau_hopping_2014}, more care is needed for systems with soft interactions. In particular, we here describe the process for a contact power-law pair interaction. The total energy $U$ is then the sum of all pair particle energies $u$ for particles $i$ and $j$, $U=\sum_{ij}u(h_{ij})$. Here, 
\begin{equation}
u(h_{ij})= \frac{\varepsilon}{\alpha}(-h_{ij})^\alpha\Theta(-h_{ij}),
\end{equation}
where $\varepsilon=1$ sets the energy scale, $\alpha>1$ the contact power, $\Theta$ is the Heaviside function, and 
\begin{equation}
h_{ij} = d\bigg(\frac{|\mathbf{r}_{i}-\mathbf{r}_j+\mathbf{\Lambda}_{ij}|}{\sigma_i + \sigma_j}-1\bigg)
\label{eq:gap}
\end{equation}
is the dimensionless gap. (Although only harmonic spheres with $\alpha=2$ are here considered, the following derivation is offered for generic $\alpha$.) Here $\mathbf{r}_{i}$ and $\sigma_i$ are the position and radius of particle $i$, respectively, and $\mathbf{\Lambda}_{ij}$ is the random offset. The normalization factor $d$ in Eq.~\eqref{eq:gap} ensures that the potential remains finite when $d\rightarrow\infty$ \cite{charbonneau_glass_2017, scalliet_marginally_2019, manacorda_gradient_2022}. For this system, the random offset $\mathbf{\Lambda}_{ij}$ needs to be generically chosen from the probability distribution
\begin{align}
P(\mathbf{\Lambda}_{ij}|\mathbf{r}_{ij}) &= \frac{\exp\big[-\beta u(|\mathbf{r}_{ij} + \mathbf{\Lambda}_{ij}|)\big]}{\int d\mathbf{\Lambda}_{ij}\exp\big(-\beta u(|\mathbf{r}_{ij} + \mathbf{\Lambda}_{ij}|)\big)} \nonumber \\
&=\frac{\exp\big[-\beta u(|\mathbf{r}_{ij} + \mathbf{\Lambda}_{ij}|)\big]}{V_\mathrm{b} -  V_d \sigma^d[1-d \int_0^1 \xi^{d-1} e^{-\frac{d^\alpha\beta}{\alpha}(1-\xi)^\alpha}d\xi]}
\label{eq:softdist}
\end{align}
at inverse temperature $\beta=\frac{1}{k_BT}$ (with Boltzmann constant $k_B$ hereon set to unity) in a simulation box of volume $V_b$, where $V_d = \frac{\pi^{d/2}}{\Gamma(1+d/2)}$ is the volume of a $d$-dimensional unit sphere. For $d^\alpha\beta/\alpha \rightarrow \infty$ the hard sphere MK model is recovered, and for $d^\alpha\beta/\alpha \rightarrow 0$ the ideal gas limit is recovered. Note that the above expressions are independent of the simulation box geometry. We here specifically use simulations boxes under $D_d$ boundary conditions (based on the $d$-dimensional generalization of the $d=3$ face centered cubic lattice) in order to most efficiently sample systems of $N$ particles~\cite{conway_fast_1982, charbonneau_dimensional_2022, ddzd2d}.

Prior simulations of the soft sphere MK model have sampled this distribution using the Metropolis-Hastings algorithm~\cite{nishikawa_relaxation_2022, nishikawa_collective_2022}, but both speed and accuracy can be improved by using inverse transform sampling (ITS)~\cite[Ch.~2.2]{devroye_nonuniform_1986}. Given the total probability that particle $i$ is contained within a distance $\sigma$ of $j$, $P(r_{ij}<\sigma)=dV_d\sigma^d\int_0^1\xi^{d-1}P(\Lambda_{ij}|r_{ij}) d\xi$, the algorithm proceeds in two steps. First, a uniform variable $u_1\in(0,1)$ is chosen. If $u_1<P(r_{ij}<\sigma)$, then simple rejection sampling can be used, wherein $\mathbf{\Lambda}_{ij}$ is sampled from the uniform distribution and rejected if the particles overlap. If $u_1\ge P(r_{ij}<\sigma)$, then $\Lambda_{ij}$ is chosen from the ITS method, for which the cumulative distribution function $\mathrm{cdf}(\Lambda_{ij}) = \int_0^{\Lambda_{ij}} P(\Lambda_{ij}|r_{ij})d\Lambda_{ij}$ is calculated numerically, and a second random variable $u_2\in(0,1)$ is chosen such that $\mathrm{cdf}^{-1}(u_2) = P(\Lambda_{ij})$. To obtain points uniformly distributed on the hypersphere, each is then multiplied by a unit vector randomly taken from a Gaussian distribution~\cite{muller_note_1959}.

\subsection{Gradient descent minimization}
\label{sec:gradDescent}

Instantaneous quenches of equilibrium configurations of soft spheres are evolved under a gradient descent (GD) algorithm. Particle motion is then overdamped, with a coefficient of static friction $\zeta$ such that
\begin{equation}
\zeta\dot{\mathbf{r}}_i = \mathbf{F}_i =  -\frac{\partial U}{\partial \mathbf{r}_i}.
\end{equation}
As such, $\mathbf{F}_i/\zeta$ has units of a velocity, and therefore
\begin{equation}
\Delta \mathbf{r}_i = \frac{\mathbf{F}_i}{\zeta} \Delta t
\end{equation}
The characteristic time $\Delta t$ is chosen such that $\max(\Delta \mathbf{r}_i)$ remains below a certain threshold fraction $\Xi$ of the diameter $\sigma$ (or the maximum overlap in the system), 
\begin{equation}
\Delta t = \frac{\zeta \Xi \sigma}{f_\mathrm{ub}},
\label{eqn:tInitial}
\end{equation}
where $f_\mathrm{ub} = \max_i \left\lVert F_i \right\rVert$ is the maximum unbalanced force on a particle. The integral equation is then
\begin{equation}
\Delta \mathbf{r}_i = \frac{\Xi \sigma}{f_\mathrm{ub}}\mathbf{F}_i.
\end{equation}

In order to minimize the computational cost while closely approximating the exact GD minimization, we use an adaptive steepest descent algorithm, refashioned from that of Refs.~\cite{folena_gradient_2021} and~\cite{stanifer_avalanche_2022}. The objective is to move slowly along the gradient direction, such that the gradient orientation only changes by a small fraction at each step. The time step $\Delta t$ is therefore chosen initially by Eq.~\eqref{eqn:tInitial}, but also follows the condition
\begin{equation}
\label{eq:forceAngle}
\sum_i^N\frac{\mathbf{F}_i(t)\cdot \mathbf{F}_i(t+\Delta t)}{\left\lVert F_i (t) \right\rVert \left\lVert F_i(t+\Delta t) \right\rVert}  = \cos\chi_{t,t+\Delta t}< 1-\epsilon
\end{equation}
with $0<\epsilon \ll 1$. If the condition is violated, the step is rejected and attempted again with $\Xi \to \Xi/n$ for $n\in\mathbb{N}$. Special care must also be taken to leave particles in zero-energy kissing contact as required by the GD dynamics. Progressively smaller steps are therefore taken upon approaching each contact breaking event, until the dimensionless overlap between particles $i$ and $j$ is $-h_{ij} < 10^{-8}$. We note that while some contact breaking events are associated with rearrangements, not all of them are~\cite{morse_differences_2020, tuckman_contact_2020}. Particles are therefore allowed to oscillate in and out of contact without rearrangement. Lowering the minimum step size in contact breaking events increases the frequency with which this happens, leading to dramatic computational slowdowns without noticeable effect on the relaxation times or the distribution of inherent states.

In order to further improve numerical efficiency, we employ a sticky algorithm which grows and shrinks the initial value of $\Xi$ according to thresholding conditions. If $n$ successive steps produce ${\cos\chi_{t,t-1}< 1-\epsilon}$ then ${\Xi \to n\Xi}$. Whereas, if $n$ steps in a row produce a $\Xi$ that must be decreased, then the initial guess of $\Xi$ becomes ${\Xi \to \Xi/n}$. Empirically, we find that $\epsilon = 10^{-2}$ and $n = 5$ give stable results for the IS energy. Because the process is chaotic~\cite{nishikawa_relaxation_2022}, changing $n$ or $\epsilon$ changes an individual IS determination, but ensemble averages remain unaffected.

A configuration minimizes either to a jammed (UNSAT) or unjammed (SAT) state, as defined by the rigidity of its contact network after (recursively) removing rattlers~\cite{morse_geometric_2014, morse_geometric_2016, artiaco_hardsphere_2022}. Although in principle it is equally valid to determine if a state is jammed by its energy, pressure, or stress being non-zero, the numerical resolution of each of these quantities is far lower than for individual contacts. One should, however, note that during the GD minimization, several contacts are transient~\cite{stanifer_avalanche_2022, manacorda_gradient_2022, cahm}, and that a change to the contact network is not always associated with entering a new energetic sub-basin~\cite{morse_differences_2020, tuckman_contact_2020}. The jamming transition $\widehat{\varphi}_J(N)$ is then determined using several systems equilibrated at a given density and temperature. For each condition, the fraction of resulting jammed states is measured. For a fixed temperature, this fraction monotonically increases with $\widehat{\varphi}$, hence $\varphi_J(N)$ is naturally defined as the density for which half of all initializations jam. Empirically, an error function is found to capture well the growth of the fraction of jammed states with density, and is therefore used as fitting form~\cite{ohern_jamming_2003}. 

\subsection{Effective potential hard sphere crunch}
\label{sec:crunch}

\begin{figure}[ht]
\includegraphics[width=\linewidth]{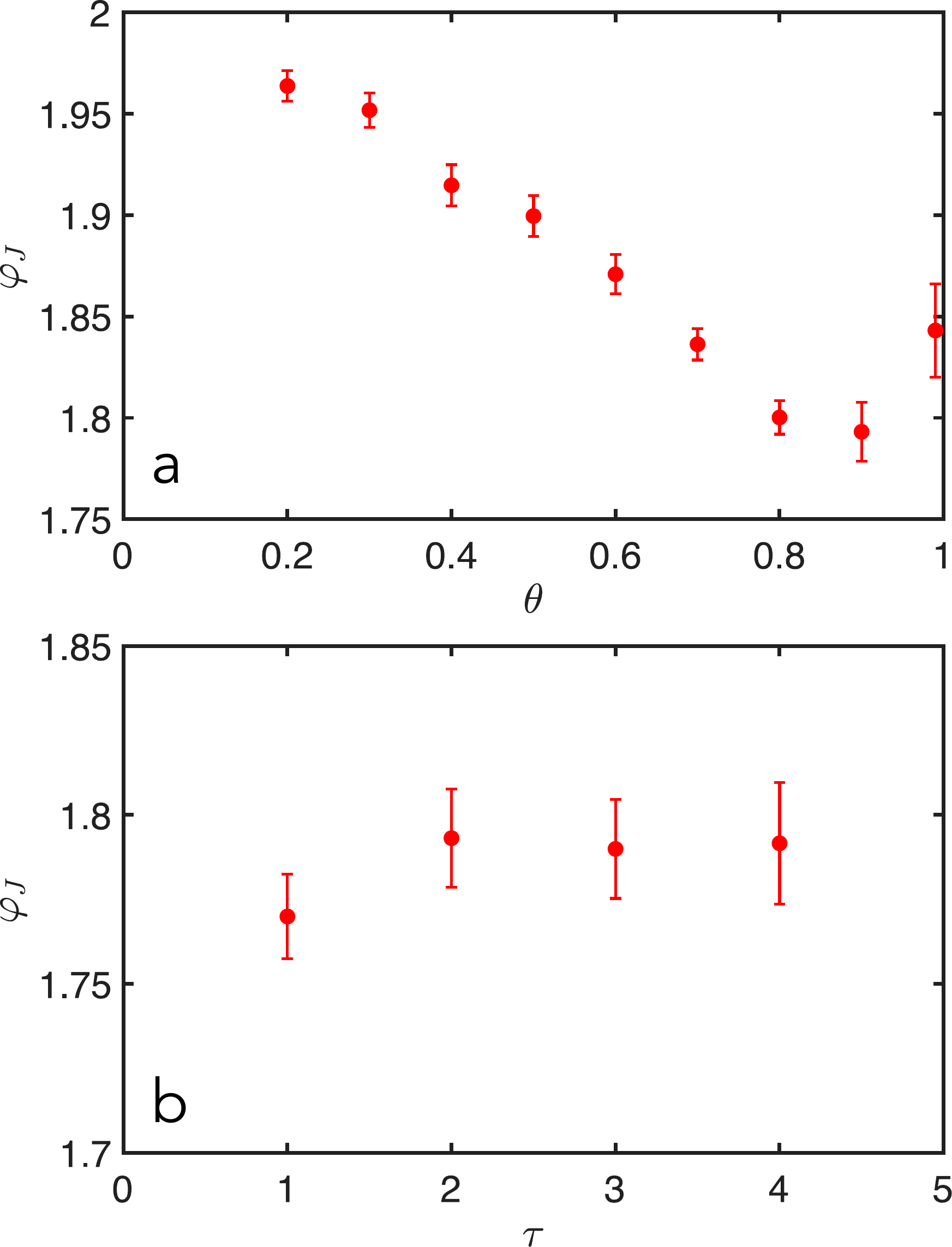}
\caption{The hard sphere crunching algorithm developed in Ref.~\cite{charbonneau_memory_2021} contains two optimization parameters, $\theta$ and $\tau$, as described in Sec.~\ref{sec:crunch}. Optimal values yield the lowest jamming density $\varphi_\mathrm{J}$ while not undergoing algorithmic arrest (leading to non-rigid packings). We here show parameter optimization for $N=1024$ particles in $d=3$ with a initial liquid density $\varphi_\mathrm{eq}=1 \ll \varphi_\mathrm{on}$, well below the onset. \textbf{a)} Using a fixed value of the cutoff $\tau=2$, the optimal expansion factor minimizes the function, $\theta \approx 0.9$. \textbf{b)} Using a fixed $\theta = 0.9$, the value of $\varphi_J$ depends only weakly on $\tau$. However, as for standard hard spheres~\cite{charbonneau_memory_2021}, $\tau=1$ corresponds to states which often lead to algorithmic arrest, and therefore do not properly jam. We use the lowest integer value for which algorithmic arrest is avoided, $\tau=2$.}
\label{fig:crunchOptimization}
\end{figure}

Hard sphere liquids are crunched following the effective potential approach (EP) of Ref.~\cite{charbonneau_memory_2021}. First, a regularization step (which is particularly important at low densities) reduces the smallest gap between particles to $1\%$ of $\sigma$ by adjusting the density accordingly. This scheme then alternatively iterates (i) an instantaneous expansion of particles, and (ii) a minimization of an effective potential via FIRE minimization~\cite{bitzek_structural_2006} to open a (normalized) gap equivalent to the (normalized) gap of the previous iteration. The process continues until an isostatic structure is formed, as determined for gaps between contacts being no greater than $10^{-8}\sigma$. The choice of effective potential is taken to be a shifted and truncated logarithm of the gaps~\cite{arceri_vibrational_2020}, which becomes the effective thermal potential for hard spheres near jamming~\cite{brito_rigidity_2006, altieri_jamming_2016}. In particular, we use $U = \sum_{ij} u(d_{ij})$ ,where $d_{ij} = -h_{ij}/d$ is the dimensionless overlap, with $h_{ij}$ given by Eq.~\eqref{eq:gap}, and  
\begin{equation}
u(d_{ij}) = \begin{cases} 
\infty& d_{ij} < 0\\
-\ln(d_{ij}) + \frac{1}{h_c}(1+d_{ij})& 0 < d_{ij} < h_c \\
0& d_{ij} \ge h_c.
\end{cases}
\end{equation}
The cutoff $h_c$ is set such that there is an average of $2d$ contacts per particle with $u(d_{ij})>0$. Two parameters are then adjusted to minimize the final jamming density: $\theta$ gives the expansion factor relative to the minimum gap in the system, and $\tau$ provides a cutoff on the minimization (and thus a cap on the effective thermalization) of the FIRE algorithm. Step (ii) terminates after $\tau N d$ minimization if it cannot be completed. For standard HS, it was found that $\tau=2$ and $\theta=0.9$ were approximately optimal. In MK HS, the same parameters appear to be near optimal, in that they minimize the jamming density obtained, as illustrated in Fig.~\ref{fig:crunchOptimization} for a typical example.

\subsection{Event-driven hard sphere crunch}
\label{sec:overdampedOld}

Hard sphere liquids are also crunched using a modified version of an event-driven (ED) crunching algorithm developed by Lerner et al.~to generalize the GD optimization scheme to hard spheres~\cite{lerner_simulations_2013}. However, unlike the affine field initially used to compress particles, we here isotropically grow particles. Although these processes are functionally similar, compression through an affine field requires manipulating the field to maintain periodic boundary conditions while isotropic growth presents no such difficulty. (Wherever possible, variable names here nevertheless follow those of Ref.~\cite{lerner_simulations_2013}.) 

ED crunches require defining a velocity on each particle, an integration scheme, and a definition of the interparticle forces. For hard particles, forces only occur at kissing, which in practice means within a small numerical error, which is here $\mathcal{O}(10^{-12}\sigma)$. By considering instantaneous changes in distances between contacting particles $i$ and $j$, with velocities $V_i$ and $V_j$ respectively
\begin{align}
\dot{r}_{ij} &= \sum_k \frac{\partial r_{ij}}{\partial \vec{r}_k} \cdot \vec{V}_k \nonumber \\
&= \sum_k (\delta_{jk} - \delta_{ik})\vec{n}_{ij} \cdot \vec{V}_k = (\vec{V}_j - \vec{V}_i) \cdot \vec{n}_{ij},
\end{align}
one can define the $\mathcal{S}$-matrix 
\begin{equation}
\label{eq:sdef}
\mathcal{S} = \frac{\partial r_{ij}}{\partial \vec{r}_{k}}.
\end{equation}
The ket notation here encodes vectors, with capitalized letters denoting the $N\times d$ dimensional space of all particles and lower-case letters denoting the $N_c$-dimensional space of contacts. From Eq.~\eqref{eq:sdef} one therefore obtains ${\ket{\dot{r}} = \mathcal{S}\ket{V}}$. In other words, the $\mathcal{S}$-matrix acts on the space of velocities to generate a space of contacts.

Through particle expansion at a rate $\dot{\gamma}$, contacting particles remain in kissing contact unless forced out by other particles, i.e.,
\begin{equation}
\label{eq:contactCondition}
\ket{\dot{r}} = \mathcal{S}\ket{V} = \ket{\dot{\sigma}} = \dot{\gamma}\ket{\sigma_0},
\end{equation}
where $\ket{\sigma_0}$ encodes the sum of radii $\sigma_{ij} = (\sigma_i + \sigma_j)/2$ between contacting particles, with values taken at time $t=0$. 

As in Ref.~\cite{lerner_simulations_2013}, we apply a simple (and standard) model for overdamped dynamics that involves embedding a system in a liquid and accounting for Stokes drag, $\ket{F^\mathrm{drag}} = -\xi^{-1}_0\ket{V}$, while ignoring other hydrodynamic forces. The contact force between kissing contacts is then simply $\ket{F^\mathrm{cont}} = \mathcal{S}^T\ket{f}$, where $\ket{f}$ encodes contact forces $f_{ij}$. Force balance dictates ${\ket{F^\mathrm{drag}} = -\ket{F^\mathrm{cont}}}$. Rearranging gives
\begin{equation}
\label{eq:lernerVF}
\ket{V} = \xi_0\mathcal{S}^T\ket{f},
\end{equation}
which can be put into the form of Eq.~\eqref{eq:contactCondition},
\begin{equation}
\mathcal{S}\ket{V} = \xi_0\mathcal{S}\mathcal{S}^T\ket{f} = \dot{\gamma}\ket{\sigma_0}.
\end{equation}
Defining $\mathcal{N} = \mathcal{SS}^T$, which is invertible unless it is singular, we can then solve for the forces
\begin{equation}
\label{eq:lernerForces}
\ket{f} = \xi_0^{-1}\dot{\gamma}\mathcal{N}^{-1}\ket{\sigma_0}
\end{equation}
Finally, inserting Eq.~\eqref{eq:lernerForces} into Eq.~\eqref{eq:lernerVF}, yields the equation of motion
\begin{equation}
\label{eq:edeom}
\ket{V} = \dot{\gamma}\mathcal{S}^T\mathcal{N}^{-1}\ket{\sigma_0}.
\end{equation}

Integration then follows the explicit formulation of Ref.~\cite{lerner_simulations_2013}, which details how contact networks are defined and updated and how the process is iterated until a jammed configuration is reached. In particular, an isostatic structure is formed with gaps between contacts which are no greater than $10^{-8}\sigma$. The only difference here is that during each integration step every particle uniformly expands its radius as ${\sigma_i \to \sigma_i(1 + \dot{\gamma}\Delta t)}$, which reduces the subsequent time to collision. For particles separated by $\Delta \vec{r}$ and time $t=0$ with relative velocity $\Delta \vec{v}$, this time of collision is given by 
\begin{equation}
\sigma^2(1 + \frac{\sigma_0}{\sigma}\dot{\gamma} t)^2 = (\Delta \vec{r} + t\Delta \vec{v})^2,
\end{equation}
which can be rewritten as
\begin{equation}
(\Delta v^2 - \sigma_0^2\dot{\gamma}^2)t^2 + 2(\Delta \vec{r} \cdot \Delta \vec{v} - \sigma\sigma_0\dot{\gamma})t + \Delta r^2 - \sigma^2 = 0.
\end{equation}
The physical solution is then
\begin{equation}
t = 
\begin{dcases}
 \frac{\Delta r^2 -\sigma^2}{2(\Delta \vec{r} \cdot \Delta \vec{v} - \sigma_0\sigma\dot{\gamma})} &\text{if } |\Delta \vec{v}| = \sigma_0\dot{\gamma} \\
\frac{\sigma_0\sigma\dot{\gamma} - \Delta \vec{r} \cdot \Delta \vec{v} - \sqrt{\kappa}}{\Delta v^2 - \sigma_0^2\dot{\gamma}^2} &\text{otherwise} \\
\end{dcases}
,
\label{eq:edIntTime}
\end{equation}
where
\begin{equation}
\kappa = (\Delta \vec{r} \cdot \Delta \vec{v})^2 - \Delta r^2 \Delta v^2 + (\sigma\Delta \vec{v} - \sigma_0\dot{\gamma}\Delta \vec{r})^2.
\end{equation}

As noted in Ref.~\cite{lerner_simulations_2013}, the natural choice of time units is $\dot{\gamma}^{-1}$ and force units is $\sigma_0\dot{\gamma}/\xi_0$. What is often less appreciated is that throughout this scheme, numerical precision and the nonlinearity of the problem introduce gaps and overlaps between contacts, particularly over large integration times (Eq.~\eqref{eq:edIntTime}). To mitigate these effects, the algorithm introduces two procedures, as detailed in the original work. First, the integration time $t_\mathrm{int}$ is taken in uniform intervals $t_\mathrm{int}\dot{\gamma}=10^{-4}$, until a collision takes place. Second, a gap correction procedure is taken periodically to either force the closure of accumulated gaps or to allow contacts to break. Taken together, these allow the systematic control of allowed overlaps, which are taken to be numerically negligible (in our case, less than $10^{-12}\sigma$).

Note that the calculation of the equation of motion Eq.~\eqref{eq:edeom} involves inverting (or solving a matrix equation of) a dense matrix of size $\mathcal{O}(dN)$ at each step, the complexity of which generically scales as $\mathcal{O}(d^3N^3)$ using optimal solvers. Additionally, the number of iterations scales as at least $\mathcal{O}(dN)$, set by the number of contacts, though in practice this estimate is a drastic underestimation. The total simulation time therefore scales at least as $\mathcal{O}(d^4N^4)$. This limitation severely constrains the accessible size and dimension range for comparing with the previous two approaches and for extrapolating to the limit $d\rightarrow\infty$.

\section{Theoretical analysis}
\label{sec:theoryHeader}

In this section, we describe the analytical estimates for the finite-$d$ dynamical transition density and the replica symmetric (RS) jamming threshold density for hard spheres under the Gaussian cage approximation. These quantities help interpret and contextualize the numerical finite-$d$ results, and their use to estimate $d\rightarrow\infty$ quantities. 

In this context, note that the mean-field $d\rightarrow\infty$ description leaves out perturbative corrections in the small parameter $1/d$~\cite{parisi_meanfield_2010}. Because the MK model separates spatial structure from other dimensional effects, leaving only the latter, characteristic densities are expected to scale linearly with that parameter, 
\begin{equation}
\label{eq:lindscalin}
\widehat{\varphi}_\mathrm{J,on,th,d}(d) = \widehat{\varphi}_\mathrm{J,on,th,d}^\infty - \tilde{a}/d +\mathcal{O}(1/d^2),
\end{equation}
where $\tilde{a}$ is a constant, and the subscripts refer to the jamming, onset, threshold, and dynamical densities, respectively. Both $\widehat{\varphi}_\mathrm{J,on,th,d}^\infty$ and $\tilde{a}$ should therefore be accessible through a linear fit from finite-$d$ results.

\subsection{Dynamical transition density}
\label{sec:phidTheory}
A finite-dimensional estimate of the mean-field dynamical transition for MK systems can be obtained under the Gaussian cage approximation as~\cite{biroli_meanfield_2021, parisi_theory_2020, biroli_liunagel_2018} 
\begin{equation}
\begin{split}
\label{eq:gca}
\frac{1}{\widehat{\varphi}_d} &= -\max_A \frac{2A}{V_d \sigma^d}\frac{\partial}{\partial A} \int q_{A/2}(\mathbf{r})\log q_{A/2}(\mathbf{r})d\mathbf{r}\\ 
&= -\max_A 2dA \frac{\partial}{\partial A} \int^\infty_1 r^{d-1} q_{A/2}(r/\sigma)\log q_{A/2}(r/\sigma) dr,
\end{split}
\end{equation}
where
\begin{equation}
\label{eq:qdef}
q_{A/2}(r) = \int^\infty_0 dy e^{-\beta u(y)}\bigg(\frac{y}{r}\bigg)^{\frac{d-1}{2}}e^{-\frac{(r-y)^2}{2A}}\frac{\sqrt{ry}}{A}I_{\frac{d-2}{2}}\bigg(\frac{ry}{A}\bigg),
\end{equation}
$u(y)$ is the interparticle potential as a function of the interparticle distance $y$, and $I_n(x)$ is the modified Bessel function. This expression can be evaluated numerically for any $d$, $u$, and $\beta$. 

For hard spheres, the potential is infinite for $0\le y< \sigma$ and zero for $y \ge \sigma$, thus resulting in the lower limit of Eq.~\eqref{eq:qdef} becoming $\sigma$ and $\exp[-\beta u(y)]\rightarrow1$. The resulting deviation of $\widehat{\varphi}_d$ from the $d\rightarrow\infty$ hard sphere result is remarkably small, even in $d=2$. Considering higher-order corrections to Eq.~\eqref{eq:lindscalin} gives $\widehat{\varphi}_\mathrm{d} = \widehat{\varphi}_\mathrm{d}^\infty - \tilde{a}/d + \tilde{a}_2/d^2 + \tilde{a}_3/d^3$, where $\widehat{\varphi}_\mathrm{d}^\infty = 4.8067\ldots$, $\tilde{a}_1 = 0.440(3)$, $\tilde{a}_2=0.38(2)$, and $\tilde{a}_3=0.91(4)$. Although non-Gaussian corrections are expected to be sizeable~\cite{biroli_local_2022}, thus possibly affecting the estimate of $\varphi_\mathrm{d}$, numerical results generally agree with this prediction~\cite{charbonneau_hopping_2014}. 

In harmonic systems, we instead analyze the dynamical temperature, $T_\mathrm{d}$, which can be obtained by numerically solving for the function $\widehat{\varphi}_\mathrm{d}(T)$, and inverting to yield $T_\mathrm{d}(\widehat{\varphi})$. Finite-$d$ corrections are similarly small.

\subsection{Finite-dimensional jamming threshold}

Using the same Gaussian-cage approximation, it is also possible to estimate the finite-$d$ correction to the RS threshold density, which has long been interpreted as the density at which jammed states appear in sufficient number for a finite complexity to be computed~\cite{parisi_meanfield_2010, monasson_structural_1995}~\cite[Sec.~7.4.3]{parisi_theory_2020}. It should be noted that the RS threshold is but an approximation of the proper $d\rightarrow\infty$ threshold, which significantly overshoots both the RLG and MK jamming transitions, thus calling into question the applicability of such calculations to this problem (see Ref.~\cite{manacorda_gradient_2022} for a thorough discussion). Nevertheless, its dimensional scaling here provides some insight into the magnitude of finite-$d$ corrections for similar quantities. 

Specifically, the approximation gives~\cite{parisi_theory_2020, mangeat_quantitative_2016}
\begin{equation}
\frac{1}{\widehat{\varphi}_\mathrm{th}} = \max_A \frac{d}{4A} \int_0^1 r^{d-1} (r-1)^2 e^{-\frac{(r-1)^2}{4A}}dr.
\end{equation}
which can also be numerically evaluated for any $d$. Fitting the asymptotic behavior, for $d\ge5$, we find $\tilde{a}=2.467(1)$, which is nearly an order of magnitude larger than for $\widehat{\varphi}_\mathrm{d}$. 

Because the RS threshold is unstable to further breaking of the replica symmetry, a one-step replica symmetry breaking (1RSB) estimate for $d\rightarrow\infty$ has also been obtained, $\widehat{\varphi}_\mathrm{th}^\mathrm{1RSB} = 6.86984\dots$~\cite{charbonneau_exact_2014}. Although this 1RSB threshold is itself unstable, higher-order of replica symmetry breaking corrections are assumed---by analogy to what is observed in certain spin glass models~\cite{rizzo_replicasymmetrybreaking_2013}---to only lightly affect its numerical value. (In any case, these corrections would only increase the resulting density.) We thus here consider $\widehat{\varphi}_\mathrm{th}^\mathrm{1RSB}$ to be a reasonable approximation of $\widehat{\varphi}_\mathrm{th}^\mathrm{fullRSB}$. Because computing $1/d$ corrections to this approximation is quite involved, however, these are left for future analysis. 

\begin{table}\centering
\caption{Parameters for the fits to Eq.~\eqref{eq:lindscalin} for the $d\to\infty$ limit ($\widehat{\varphi}^\infty$) and the scaling prefactor with $1/d$ ($\tilde{a}$). RS and 1RSB calculations of $\widehat{\varphi}^\infty$ are exact and are reported here with the highest published precision~\cite{parisi_theory_2020, charbonneau_exact_2014}; finite-$d$ corrections to the 1RSB calculation have not been computed; all other quantities are fitted to data, with parentheses giving the 95\% confidence interval}
\begin{tabular}[t]{|c | c | c | c |}
\hline
Quantity & RS/1RSB & $\widehat{\varphi}^\infty$ & $\tilde{a}$\\
\hline
$\widehat{\varphi}_\mathrm{d}$ & RS & 4.8067\dots & 0.440(3) \\
\hline
$\widehat{\varphi}_\mathrm{th}$ & RS & 6.2581\dots & 2.467(1) \\
\hline
$\widehat{\varphi}_\mathrm{th}$ & 1RSB & 6.86984\dots & -- \\
\hline
\end{tabular}
\label{table:phiOthersCompare}
\end{table}

\section{Results and Discussion}
The numerical results for relaxation, crunching, and IS memory in MK systems obtained as described in Sec.~\ref{sec:methods} are here compared with the analytical results from Sec.~\ref{sec:theoryHeader} and the RLG results of Ref.~\cite{manacorda_gradient_2022}.

\begin{figure*}[ht]
\includegraphics[width=0.9\linewidth]{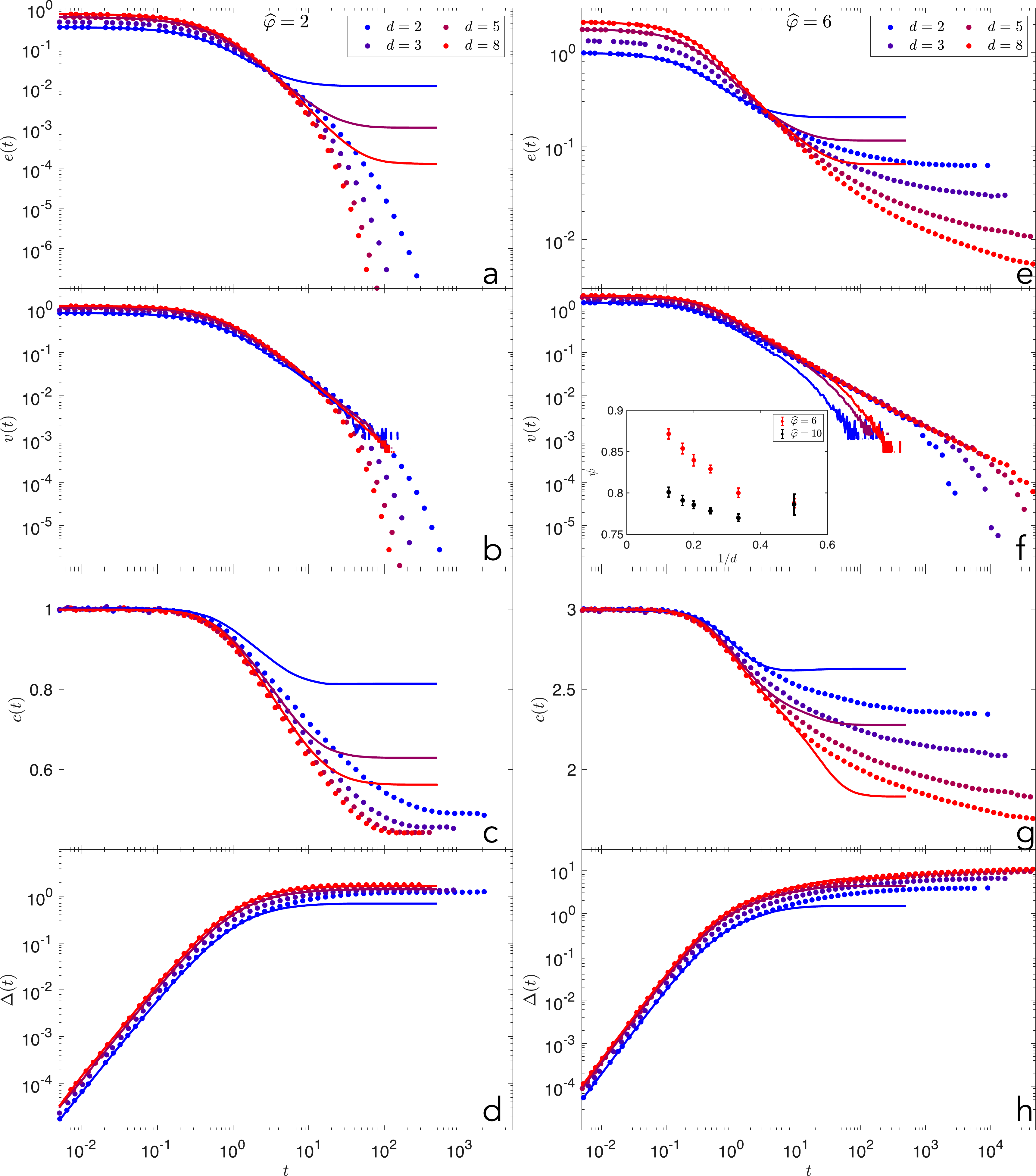}
\caption{Relaxation dynamics across dimension both (a-d) below jamming (at $\widehat{\varphi}=2$) and (e-h) above jamming (at $\widehat{\varphi}=6$) of the energy (a,e) $e(t)$, (b,f) velocity $v(t)$, (c,g) scaled contact number $c(t)$, and (d,h) MSD $\Delta(t)$. The MK results (points) in $d=2, 3, 5,$ and $8$ are compared with RLG results (lines) in $d=2, 5,$ and $8$ taken from Ref.~\cite{manacorda_gradient_2022} (after rescaling to make the DMFT descriptions equivalent, as in Eq.~\eqref{eqn:RLGCorr}). The inset in (f) shows the power-law scaling behavior given by Eq.~\eqref{eq:relaxationVelocity} as a function of $d$ for two densities above jamming, $\widehat{\varphi}=6$ and $\widehat{\varphi}=10$.}
\label{fig:relaxation}
\end{figure*}

\begin{figure*}[ht]
\includegraphics[width=\linewidth]{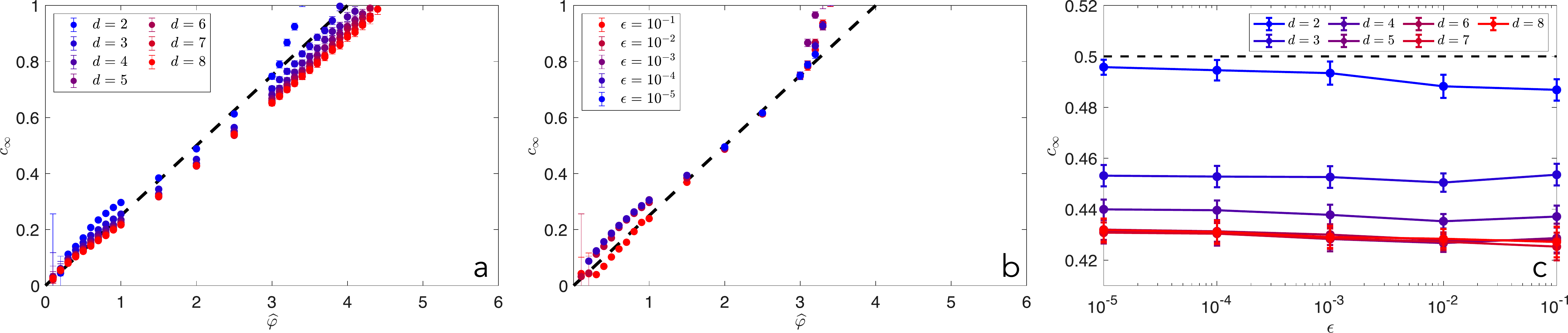}
\caption{(a) Final contact number, $c_\infty$ in systems below jamming for various $d$ and $\widehat{\varphi}$ for $\epsilon=10^{-2}$. The low density asymptotic prediction $c_\infty \approx \widehat{\varphi}_\mathrm{MK}/4$ (dashed black line) is approached as $d\to\infty$ for asymptotically low density.  (b) Same quantity in $d=2$ for several $\epsilon$, demonstrating that $c_\infty$ rapidly converges at all densities for $\epsilon < 10^{-1}$. (c) At $\widehat{\varphi}=2$, $c_\infty$ shows a much stronger dependence on $d$ than on $\epsilon$.}
\label{fig:cInf}
\end{figure*}

\subsection{Relaxation dynamics}
\label{sec:relaxation}

We first consider GD dynamics of MK soft spheres, starting from an infinite-temperature configuration with purely random initial particle positions. By examining the relaxation, aging, and the unstressed (SAT) or frustrated (UNSAT) nature of the final state, we seek to validate comparable observations for the RLG~\cite{manacorda_gradient_2022}. We specifically compare and contrast the behavior of the two models far below ($\widehat{\varphi}_\mathrm{MK}=2$) and far above ($\widehat{\varphi}_\mathrm{MK}=6$) jamming, after rescaling density, mean squared displacement $\Delta$ (MSD), and characteristic timescale as~\cite{manacorda_gradient_2022}
\begin{equation}
t_\mathrm{MK} = 2t_\mathrm{RLG}, \ \ \Delta_\mathrm{MK} = \frac{1}{2}\Delta_\mathrm{RLG}, \ \ \widehat{\varphi}_\mathrm{MK} = 2\widehat{\varphi}_\mathrm{RLG},
\label{eqn:RLGCorr}
\end{equation}
such that the two models should coincide in the $d\rightarrow\infty$ limit. Because the MK and RLG models present distinct numerical and theoretical challenges in finite $d$, features that robustly appear in both almost surely also appear in the DMFT.

Figure~\ref{fig:relaxation} shows that the initial time relaxation of the MK and RLG models are essentially equivalent in all $d$, both below and above jamming. This correspondence reflects the single-particle nature of optimization at very short times. For $t \ge 1$, the many-body nature of the MK model results in quantitative dynamical differences, but as $d$ increases, this difference seemingly vanishes as well. In particular, for both models energy and velocity decay exponentially below and algebraically above jamming~\cite{nishikawa_relaxation_2022, manacorda_gradient_2022},
\begin{equation}
\label{eq:relaxationVelocity}
\langle |\mathbf{v}(t)| \rangle \sim t^{-\psi}; \ \ \langle E(t) - E(t\rightarrow\infty)\rangle \sim t^{-(2\psi-1)}.
\end{equation} 
We find that the weak dimensional dependence of $\psi$ for two such densities (Fig.~\ref{fig:relaxation}f inset) is numerically consistent with previous reports for RLG and $d=3$ inverse-power MK models~\cite{manacorda_gradient_2022, nishikawa_relaxation_2022}. Furthermore, only minor deviations are observed in a variety of standard non-MK interparticle potentials (including both $d=2$ and $d=3$ Kob-Andersen Lennard-Jones and inverse-power potentials and $d=2$-$4$ and $d=8$ harmonic potentials)~\cite{nishikawa_relaxation_2022}. This behavior should therefore be considered to be a robust feature of the DMFT.
 
The MSD of the MK model likewise converges to the RLG behavior as $d\rightarrow\infty$. Below jamming, $\Delta$ quickly plateaus as a stable unjammed state is reached. Above jamming, by contrast, the MSD continuously drifts upward. This drifting regime coincides with the algebraic decay of the energy and velocity, thus indicating that the system reaches ever lower energy states. While the drift is partially obscured in the low-$d$ RLG, in the MK model it is clearly visible even in $d=2$. We therefore confirm that whether a liquid jams or not following GD minimization results in two drastically different behaviors. The former does so exponentially quickly; the latter slowly finds ever lower energy states in a process of athermal aging, as previously reported in Refs.~\cite{nishikawa_relaxation_2022, nishikawa_relaxation_2021, olsson_relaxation_2022, parley_meanfield_2022, manacorda_gradient_2022}. Within a context of finite-precision calculations, the optimization process is necessarily truncated (e.g., once the gradient falls below a given threshold). This robustly universal scenario therefore challenges the traditional assumption that jammed states obtained by simple minimization are purely static~\cite{ohern_jamming_2003}. This effect, while never reported in non-MK systems, is nevertheless expected to affect generic soft sphere systems, as evidenced by the power-law behavior of the velocity in Ref.~\cite{nishikawa_relaxation_2022}. Confirmation of this effect, however, is left for future work.

A key discrepancy between the RLG and MK results is nevertheless observed in the average contact number at long times, $c_\infty$. In both systems, low density states are expected to approach $c_\infty \equiv \lim_{t\to\infty} c(t) \approx \widehat{\varphi}_\mathrm{MK}/4$ in the joint dilute, $\widehat{\varphi} \ll\widehat{\varphi}_J$, and high-dimensional limits~\cite{manacorda_gradient_2022}. In the RLG, this relationship holds until $\widehat{\varphi}_\mathrm{RLG}\approx1.25$ ($\widehat{\varphi}_\mathrm{MK}\approx2.5$), and is therefore captured in Fig.~\ref{fig:relaxation}c, for which $\widehat{\varphi}_\mathrm{RLG}=1$. The curves, however, differ significantly, with $c_\infty^\mathrm{RLG}\to0.5$ and $c_\infty^\mathrm{MK}\to0.425$. (See Ref.~\cite{manacorda_gradient_2022} for higher-dimensional data and a discussion of the $d\to\infty$ limit.) To make sense of this difference, we consider two possible sources of discrepancy: the density and $\epsilon$ dependence of $c_\infty^\mathrm{MK}$.

For $\widehat{\varphi}_\mathrm{MK} \le 0.5$, the relationship $c_\infty \approx \widehat{\varphi}_\mathrm{MK}/4$ appears to hold as $d\to\infty$ (Fig.~\ref{fig:cInf}a), in agreement with the analytical result for asymptotically low densities. However, significant deviations develop beyond that point, which is far below where they emerge for the RLG. Could it be that the multi-body nature of the MK model somehow reduces the quality of the minimization compared to the single-particle RLG model? It seems not. Decreasing the size of the smallest time step, $\epsilon$, by orders of magnitude results in only minuscule quantitative changes compared to $\epsilon=10^{-2}$---the value used in the rest of this work. Figure~\ref{fig:cInf}b presents a case study in $d=2$, and Fig.~\ref{fig:cInf}c shows that the effect only weakens as $d$ increases. The difference between the MK and RLG contact numbers therefore persists in the slow minimization, $\epsilon \to 0$, and thermodynamic, $N\to\infty$, limits for any finite $d$. Given the theoretical expectation that the two should coincide in the $d\to\infty$ limit, this difference might be due to (stronger than expected) pre-asymptotic corrections in the RLG, thus obfuscating the dimensional extrapolation. We note, in particular, that finite-$d$ RLG minimizations below jamming do not systematically result in configurations with zero energy, unlike comparable MK minimizations.

Whatever the physical origin of this result, it has marked consequences on the extraction of $\widehat{\varphi}_J$, which corresponds to $c_\infty = 1$ in these units. The pronounced deviation of the MK results from the low density scaling $c_\infty \approx \varphi_{MK}/4$ results in a higher jamming density than for the RLG, as can be qualitatively observed in Fig.~\ref{fig:cInf}.  In order to quantitatively estimate $\widehat{\varphi}_J$, in Sec.~\ref{sec:hardsoft} we follow the more computationally efficient approach described in Sec.~\ref{sec:gradDescent} using the fraction of systems jammed as a function of $\widehat{\varphi}$. The difference between the approaches here and there amounts to a swapping of the high-dimensional and thermodynamic limits, which is not here expected to result in any numerical difference, as long as the fraction of rattlers, which decays exponentially quickly with dimension~\cite{charbonneau_jamming_2015}, is sufficiently suppressed.

\begin{figure}[ht!]
\includegraphics[width=0.9\linewidth]{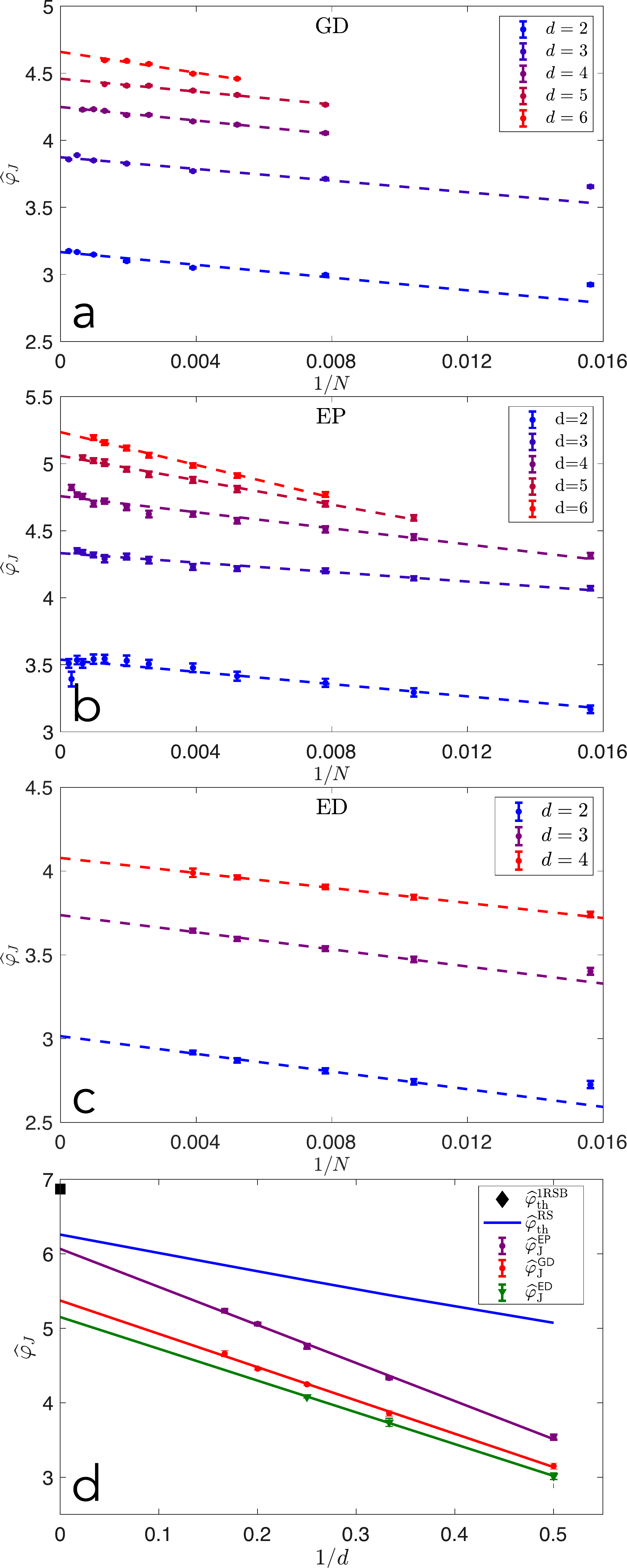}
\caption{Finite-size scaling of the jamming transition using (a) gradient descent (GD) minimization as well as (b) effective potential (EP) and (c) event-driven (ED) crunches. For each, linear fits to Eq.~\eqref{eq:phiJN} are given. (d) Comparison of the $N\to\infty$ jamming transition and threshold densities in MK systems with linear fits to Eq.~\eqref{eq:lindscalin}; see Table~\ref{table:phiJCompare} for parameters.}
\label{fig:phiJN}
\end{figure}

\subsection{Hard and Soft Sphere Jamming}
\label{sec:hardsoft}
In order to compare jamming density results with the RLG and DMFT analysis, the lowest jamming density achieved using the different schemes is considered. For soft spheres, configurations are first equilibrated at $T=\infty$, which corresponds to the canonical quench of Refs.~\cite{ohern_jamming_2003, morse_geometric_2014, manacorda_gradient_2022}. For hard spheres, all liquid configurations equilibrated below an onset density $\widehat{\varphi}_\mathrm{on}$ crunch towards the same (lowest) jamming density. 

To extract the thermodynamic jamming transition in a given dimension, $\widehat{\varphi}_J(d,\infty)$, we use as scaling form (see Fig.~\ref{fig:phiJN}) 
\begin{equation}
\label{eq:phiJN}
\widehat{\varphi}_\mathrm{J}(d,\infty) - \widehat{\varphi}_\mathrm{J}(d,N) \sim 1/N.
\end{equation}
which satisfactorily describes the finite-size scaling for all $d$ and algorithms considered. Equation~\eqref{eq:phiJN}, however, contrasts strongly with the scaling form used for non-MK systems, which follows ${\widehat{\varphi}_J(d,\infty)-\widehat{\varphi}_J(d,N)\sim N^{-\nu/d}}$ with $\nu\lesssim 1$~\cite{ohern_jamming_2003, charbonneau_memory_2021, vagberg_finitesize_2011, olsson_dynamic_2020}. Given that the upper critical dimension for jamming has been argued to be $d_u=2$, one might have expected finite-size corrections for both MK and non-MK models to scale similarly in $d>2$. It is unclear why they don't. It is nevertheless interesting to note that the form of Eq.~\eqref{eq:phiJN} is consistent with that predicted by Wittmann and Young for $d>d_u$~\cite{wittmann_finitesize_2014}.

In all three cases the dimensional scaling appears to follow Eq.~\eqref{eq:lindscalin} (see Fig.~\ref{fig:phiJN}d and Table~\ref{table:phiJCompare}), with a roughly similar dimensional prefactor. Just like for the Gaussian cage approximation discussed in Sec.~\ref{sec:phidTheory}, that scaling prefactor is about an order of magnitude larger than for $\widehat{\varphi}_\mathrm{d}$. The (lack of) dimensional dependence in $\widehat{\varphi}_\mathrm{d}$ might therefore reflect a cancellation of terms in Eq.~\eqref{eq:gca}. (The physical origina of that cancellation, however, is not obvious at this point.) More significantly, all three estimates for $\widehat{\varphi}_J(d\rightarrow\infty)$ are much lower than $\widehat{\varphi}_\mathrm{th}$, thus confirming the distinctness of that theoretical quantity from actual jamming densities. 

\begin{figure*}[!htpb]
\includegraphics[width=\linewidth]{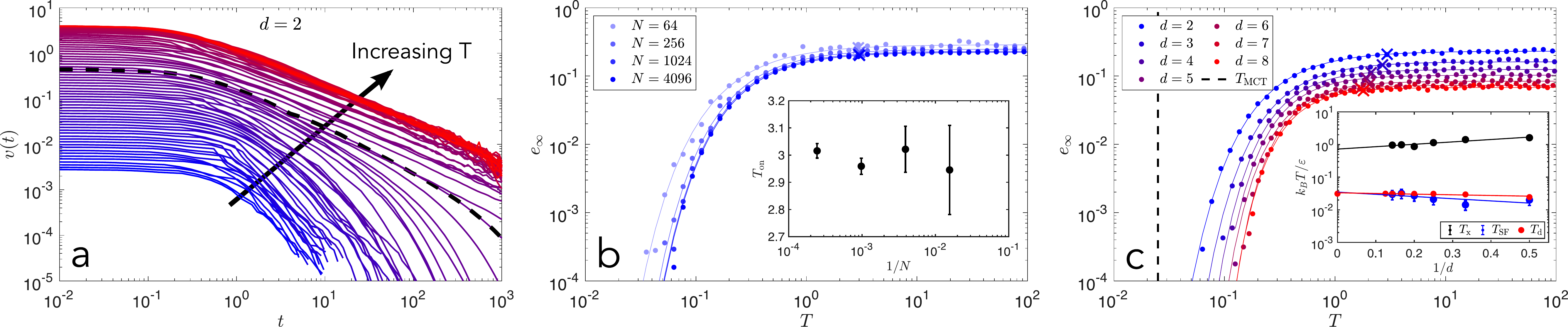}
\caption{(a) Velocity evolution of $d=2$ soft spheres undergoing a GD minimization from configurations initially equilibrated at $T=10^{-3}$ to $10^2$. The change from exponential to algebraic decay is fairly sharp (dashed line), thus identifying $T_\mathrm{SF}$.  (b) The inherent state energy for $d=2$ over the same temperature is found to vary only weakly with system size. (inset) The crossover $T_\mathrm{x}$ (crosses), as defined in the text, is similarly only weakly $d$ dependent. (c) The inherent state energy in $d=2$-$8$ presents a smooth $d$ dependence.  (inset) The weak dimensional trend of both $T_\mathrm{x}$ and $T_\mathrm{SF}$ are consistent with $T_\mathrm{SF} \le T_\mathrm{d} < T_\mathrm{x}$. As guide to the eye, in (b) and (c) the inherent state energy are phenomenologically described to $\log e_\infty = a-bT^{-c}$ where $a$, $b$, and $c$ are fit parameters.}
\label{fig:softOnset}
\end{figure*}

The numerical results for the different algorithms, however, are not consistent with each other. In particular, the EP scheme gives much larger jamming densities than soft sphere GD and hard sphere ED. The difference might be due to EP minimization permitting some degree of (effective) thermalization compared to the other two. As further support for this interpretation, we note that a previous soft sphere estimate obtained using a (FIRE) minimizer---known for having a higher degree of thermalization than GD---also resulted in significantly denser jammed states~\cite{parisi_theory_2020}. Whatever the physical origin of this effect, it is much more pronounced in MK than in non-MK model spheres~\cite{charbonneau_comment_2022}. This marked difference likely reflects the lack of a well-defined liquid shell structure to accompany caging in MK models, thus expanding the number of possible relaxation pathways. Although this effect has not yet been systematically studied--and is left as future work--this analysis already makes it clear that the EP scheme is generally ill-suited for identifying low-density jammed states. (The recently described CALiPPSO scheme is expected to fare even worse by this metric, given its built-in reliance on thermalization~\cite{artiaco_hardsphere_2022}.)
 
One might posit that the difference between HS crunches and SS minimization originates from hard and soft spheres experiencing significantly different landscapes, with some of the barriers in the former turning into mere saddles in the latter, for example. The ED crunching scheme results support this interpretation, albeit only marginally. In low $d$, ED--the GD analog for HS--reaches jamming at slightly lower densities than actual GD for SS. The extrapolation error, however, is too large to discern whether the effect is systematic or not in the limit $d\rightarrow\infty$. In any event, this near coincidence between two markedly different algorithms (see Sec.~\ref{sec:relaxation}) suggests that the $d\rightarrow\infty$ jamming density estimate they provide might be closer to the relevant DMFT prediction than that obtained from the RLG model.

\begin{table}\centering
\caption{Summary of the the $d\to\infty$ limit of $\widehat{\varphi}_J$ as extracted from results from GD minimization of the RLG, GD and FIRE minimization of the MK, as well as the effective potential (EP) and event-driven (ED) hard sphere crunches}
\begin{tabular}[t]{|c | c | c | c | c |}
\hline
$\widehat{\varphi}_J$ & $\tilde{a}$ & System & Minimizer & Ref.\\
\hline
4.2(1) & -- & RLG & GD & ~\cite{manacorda_gradient_2022}\\
\hline
5.8 & -- & MK & FIRE & ~\cite[Fig.~9.2]{parisi_theory_2020}\\
\hline
5.37(10) & 4.5(3) & MK & GD & this work \\
\hline
6.08(14) & 5.3(5) & MK & EP & this work\\
\hline
5.1(3) & 4.3(7) & MK & ED & this work \\
\hline
\end{tabular}

\label{table:phiJCompare}
\end{table}

\subsection{Inherent state memory}
\label{sec:memory}
Section~\ref{sec:hardsoft} considered infinite temperature quenches for soft spheres and low density crunches for hard spheres, aiming to achieve low density jammed systems. We now consider the impact of changing the equilibration temperature and density, respectively, on IS determination.

By analogy with abstract models of spin glasses, the GD minimization of soft spheres is expected to exhibit two characteristic temperatures: a state following temperature $T_\mathrm{SF}$ and an onset temperature $T_\mathrm{on}$, with $T_\mathrm{SF} < T_\mathrm{d} < T_\mathrm{on}$~\cite{folena_rethinking_2020}. The first separates two relaxation regimes. For $T \ge T_\mathrm{SF}$, the relaxation to the nearest minimum follows a non-trivial power law, whereas for $T <  T_\mathrm{SF}$, the relaxation to the nearest minimum is exponentially fast. The second is such that systems equilibrated at $T>T_\mathrm{on}$ (or $\widehat{\varphi}_\mathrm{eq} < \widehat{\varphi}_\mathrm{on}$)
minimize to a statistically similar set of inherent states, whereas those equilibrated at $T<T_\mathrm{on}$ (or $\widehat{\varphi}_\mathrm{eq} > \widehat{\varphi}_\mathrm{on}$) result in IS that depend on $T$~\cite{sastry_signatures_1998, sastry_onset_2000, kamath_thermodynamic_2001, brumer_meanfield_2004, sastry_glassforming_2013} (or $\widehat{\varphi}_\mathrm{eq}$~\cite{ozawa_jamming_2012, berthier_equilibrium_2016, ozawa_exploring_2017, jin_jamming_2021, charbonneau_memory_2021}). For crunching algorithms, a similar onset density is expected, but the minimization dynamics does not naturally lend itself to a state-following analysis.

Results for the GD minimization of soft spheres for $d=2$-$8$  are presented in Fig.~\ref{fig:softOnset}. The systematic study of $d=2$ systems suggests that finite-size effects are fairly weak. Relatively small system sizes, $N=1024$ for $d=2$-$5$ and $N=2048$ for $d=6$-$8$ (Fig.~\ref{fig:softOnset}c), are therefore considered. Strictly speaking, however, the value of $T_\mathrm{SF}$ extracted is an upper bound, as finite-size systems give rise to an exponential cutoff to an otherwise power-law decay. The velocity decay results scale similarly in all dimensions, thus allowing $T_\mathrm{SF}$ to be straightforwardly extracted (Fig.~\ref{fig:softOnset}c inset). This dimensional trend validates the robustness of the power-law relaxation regime previously reported for wide variety of glass formers--including $d=3$ MK~\cite{nishikawa_relaxation_2021, nishikawa_relaxation_2022}. The effect should therefore persist in the $d\rightarrow\infty$ limit, and thus be described by the DMFT.

The onset temperature, however, is more problematic. Because the IS energy varies for all $T$, no temperature cleanly separates a $T$-independent regime in any finite $d$. In soft spheres, the onset is therefore seemingly not a true transition, in agreement with recent results from spin glass models which strongly suggest that $T_\mathrm{on}=\infty$~\cite{folena_rethinking_2020,folena_weak_2023}. In order to characterize the dimensional evolution of inherent state energy, we nevertheless define a crossover temperature such that, $e(T_\mathrm{x}) = f e_\infty$ with arbitrary parameter $f$.  Here, we choose $f=0.9$, but the dimensional trend is robust to even fairly large changes to $f$. Empirically, we find that the logarithm of the resulting characteristic temperatures scale linearly with $1/d$,
\begin{equation}
\log T(d) = \log T^\infty + \tilde{b}/d,
\end{equation}
with $T_\mathrm{x}^\infty = 0.73(18)$ with $\tilde{b} = 1.7(9)$ and $T_\mathrm{SF}^\infty = 0.035(17)$ with $\tilde{b} = -2(2)$. Using the results for $T_\mathrm{d}$ from Sec.~\ref{sec:phidTheory}, we therefore confirm that $T_\mathrm{SF} \le T_\mathrm{MCT} < T_\mathrm{x}$ as expected, though we cannot conclude whether $T_\mathrm{SF}$ and $T_\mathrm{d}$ are distinct or not in the $d\to\infty$ limit. If a difference does persist in the DMFT, however, it would likely have to be quite small.

\begin{figure}[!htp]
\includegraphics[width=0.9\linewidth]{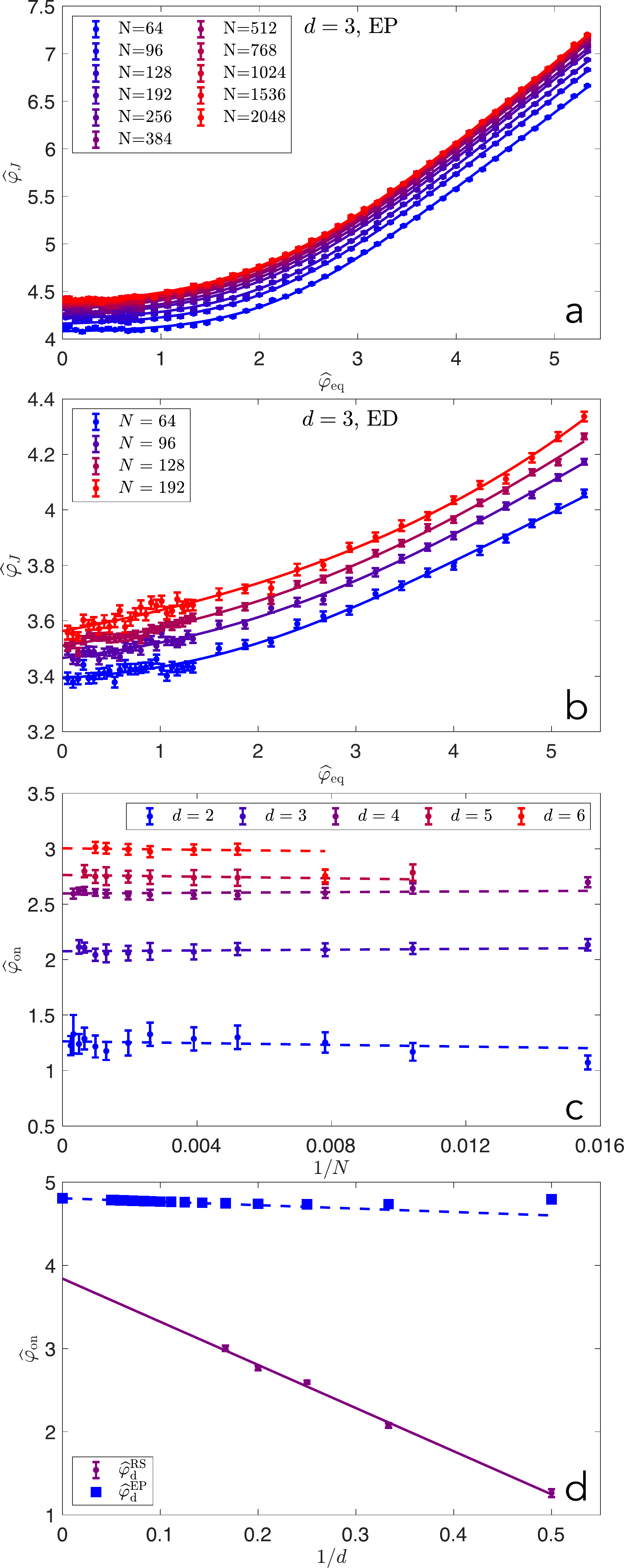}
\caption{(a) EP and (b) ED crunching exhibit strong signatures of inherent state memory, wherein $\widehat{\varphi}_J$ is a function of the initial equilibrated density $\widehat{\varphi}_\mathrm{eq}$. Data is shown here for a variety of system sizes in $d=3$, fitted to Eq.~\eqref{eq:epOnset} (solid lines). (c) From the fit in (a), the onset density $\widehat{\varphi}_\mathrm{on}$ is found to have only a weak $N$ dependence. The value extracted for $\widehat{\varphi}_\mathrm{J0}$ is shown in Fig.~\ref{fig:phiJN}d. (d) Comparison of the hard sphere MK dynamical transition from Eq.~\eqref{eq:gca} and the onset density for the  EP protocol from Eq.~\eqref{eq:epOnset}. Linear fits to Eq.~\eqref{eq:lindscalin} (lines) suggest that the two remain distinct as $d\to\infty$.}
\label{fig:epOnset}
\end{figure}

The HS crunching algorithms are more consistent with the presence of a non-trivial onset, although $\widehat{\varphi}_\mathrm{on}=0$ cannot be excluded either (see Fig.~\ref{fig:epOnset}). The IS density indeed seemingly follows the same empirical form as standard HS~\cite{charbonneau_memory_2021}, 
\begin{equation}
\label{eq:epOnset}
\widehat{\varphi}_\mathrm{J}(\widehat{\varphi}_\mathrm{eq}) = \widehat{\varphi}_\mathrm{J0} + ab\log\big(1+\exp[(\widehat{\varphi}_\mathrm{eq}-\widehat{\varphi}_\mathrm{on})/b]\big),
\end{equation}
where $\widehat{\varphi}_\mathrm{J0}$ is the lowest jamming density quoted in Sec.~\ref{sec:hardsoft}, and $a$ and $b$ are fit parameters.  Following Eq.~\eqref{eq:phiJN} (and noting only a light system size dependence), we can extract a thermodynamic estimate of $\widehat{\varphi}_\mathrm{on}$ for each $d$ (Fig.~\ref{fig:epOnset}b) on the EP data. The resulting estimate is well described by Eq.~\eqref{eq:lindscalin} (Fig.~\ref{fig:epOnset}d), thus yielding $\widehat{\varphi}^\infty_\mathrm{on}=3.84(18)$ with $\tilde{a} = 5.2(6)$. As expected, this value is well below $\widehat{\varphi}_\mathrm{d}$ (see Sec.~\ref{sec:phidTheory}) as for standard hard spheres. The scaling form used, however, is equally compatible with  $\widehat{\varphi}_\mathrm{on}$ being a crossover, and therefore small deviations could well persist even in the low-density limit.

Because the ED crunches are so computationally costly, we are here only able to demonstrate that $\widehat{\varphi}_\mathrm{on}$ exists for small system sizes. This observation nevertheless confirms that an onset $\widehat{\varphi}_\mathrm{on}$ is present in both crunching schemes considered here, thus complementing the fast compression results for $d=3$ reported in Ref.~\cite{mari_dynamical_2011}. We therefore expect the DMFT should to predict such an onset, distinct from $\phi_\mathrm{d}$, be it trivial or not. 

\section{Conclusion}
\label{sec:conclusion}

The MK model lacks spatial structure, thus making it ideal for anticipating the DMFT behavior through $d\to\infty$ extrapolations. In particular, its finite-$d$ corrections to $\widehat{\varphi}_\mathrm{d}$ and $\widehat{\varphi}_J$ are linear in $1/d$ (Eq.~\eqref{eq:lindscalin}), which contrast sharply with the highly non-trivial $1/d$ dependence of comparable quantities in standard hard sphere~\cite{charbonneau_comment_2022, charbonneau_dimensional_2022}. We have here shown that the MK model also appears to have a distinct advantage over the RLG, in that the former exhibits finite-dimensional corrections that converge faster than for the latter as $d\to\infty$. In addition to providing stronger quantitative estimates of $\widehat{\varphi}_J$, we have shown that the MK model is well suited to address the existence of the onset and its persistent differentiation from the dynamical transition in all dimensions. Because as yet no solutions to the DMFT exist, and because the MK and RLG models are expected to converge in the $d\to\infty$ limit, it nevertheless behooves us to use whichever model is best suited to the DMFT quantity of interest, while understanding that the differences between these models may also provide an understanding of generic DMFT solutions. 

Several questions that target the notion of jamming more broadly also remain. First, both MK and RLG models exhibit a logarithmic growth of the MSD (Sec.~\ref{sec:relaxation}), upon GD quenches above jamming even in finite dimensions, thus calling into question the notion that jamming is purely static. Yet, to the best of our knowledge, this feature has not been reported in standard soft sphere models. Second, while the power-law relaxation of the velocity in systems prepared above jamming has been shown to be a generic landscape feature of glasses~\cite{nishikawa_relaxation_2022, manacorda_gradient_2022}, the universality of the associated exponent has not been explained, nor has its clear dependence on $\widehat{\varphi} - \widehat{\varphi}_J$ in systems which exhibit jamming (Fig.~\ref{fig:relaxation}f inset). 
Third, while near optimal in non-shifted hard sphere models, the crunching algorithm via the log potential here far overshoots the jamming density found through soft sphere relaxation in all dimensions, making this model clearly suboptimal. ED crunches fare far better in this respect, achieving a jamming density less than that of the soft sphere MK following gradient descent, albeit at a very steep computational cost. A clearer understanding of the algorithmic difficulty involved in identifying (low-density) jammed hard spheres through crunching might finally offer some well-needed insight into the classic experiments reporting \emph{random close packing} densities in various systems.

\begin{acknowledgements}
We thank Giampaolo Folena, Alessandro Manacorda and Francesco Zamponi for useful discussions and the latter two also for sharing their RLG data. This work was supported by a grant from the Simons Foundation (Grant No.~454937). The simulations were performed at both Duke Compute Cluster (DCC)---for which the authors thank Tom Milledge’s assistance---and on Extreme Science and Engineering Discovery Environment (XSEDE), which is supported by National Science Foundation grant number ACI-1548562. Data relevant to this work have been archived and can be accessed at the Duke Digital Repository~\cite{data}.
\end{acknowledgements}

\bibliography{onset,footnotes}

\begin{thebibliography}{89}
\bibitem [{ddz()}]{ddzd2d}%
  \BibitemOpen
  \href@noop {} {}\bibinfo {note} {For $d=2$, the $D_d$ boundary conditions
  reduce to that of $Z_d$, the standard square lattice.}\BibitemShut {Stop}%
\end{thebibliography}

\end{document}